\documentclass{mem}
\usepackage{natbib}\usepackage{txfonts}\usepackage{balance}
\usepackage{graphicx}
\usepackage[a4paper]{hyperref}

\newcommand{\myemail}{viggoh@astro.uio.no}

\newcommand{\HeII}{He~II}
\newcommand{\CIV}{C~IV}
\newcommand{\OVI}{O~VI}
\newcommand{\SiVII}{Si~VII}
\newcommand{\FeXII}{Fe~XII}
\newcommand{\FeXV}{Fe~XV}
\newcommand{\aver}[1]{\langle{#1}\rangle}

\idline{1}{1}
\begin{document}
\def\teff{$T\rm_{eff }$}
\def\kms{$\mathrm {km s}^{-1}$}
\def\labs{\mid\!}
\def\rabs{\!\mid}

\title{
On red-shifts in the Transition Region and Corona
}

   \subtitle{}

\author{V.H. Hansteen\inst{1,2,3,4}
\and
H. Hara\inst{2}
\and
B. de Pontieu\inst{3}
\and
M. Carlsson\inst{1,4}
}

\institute{Institute of
 Theoretical Astrophysics, University of Oslo, P.O. Box 1029
 Blindern, N-0315 Oslo, Norway
\and
National Astronomical Observatory of Japan,
 2---21--1 Oswawa, Mitaka, Tokyo 181--8588, Japan
\and
Lockheed Martin Solar \& Astrophysics Lab, Org.\ ADBS,
 Bldg.\ 252, 3251 Hanover Street Palo Alto, CA~94304 USA
\and
Center of Mathematics for Applications,
  University of Oslo, P.O. Box 1053
  Blindern, N-0316 Oslo, Norway
\email{\myemail}}

  \offprints{V.H. Hansteen}

\authorrunning{Hansteen et al. }

\titlerunning{Red-Shifts in the TR}

\abstract{
We present evidence that transition region red-shifts are naturally produced in episodically heated models where the average volumetric heating scale height lies between that of the chromospheric pressure scale height of 200~km and the coronal scale height of 50~Mm. In order to do so we present results from 3d MHD models spanning the upper convection zone up to the corona, 15~Mm above the photosphere. Transition region and coronal heating in these models is due both the stressing of the magnetic field by photospheric and convection `zone dynamics, but also in some
models by the injection of emerging magnetic flux.
}
\maketitle{}

\section{Introduction}

Observations of the upper chromosphere and of the transition region point to a number of mysteries that have yet to be adequately explained. 

Chief among these are the rise in the differential emission measure (DEM) at temperatures below $10^5$~K \citep{1992str..book.....M} and the ubiquitous average red-shift found in transition region lines 
\citep{1998ApJS..114..151C,1999ApJ...522.1148P}           
In short, the rise in the DEM at lower transition region temperatures indicates that much more plasma is emitting in this temperature range than is accounted for by models where transition region structure is dominated by thermal conduction from overlying coronal structures. The average red-shift has been explained by a variety of models, {\it e.g.} 
\citet{1982ApJ...255..743A} 
and 
\citet{1984ApJ...287..412A} 
conjectured that red-shifts were due the return of previously heated spicular material, 
on the other hand \citet{1993ApJ...402..741H} 
argued that nanoflare induced waves generated in the corona could produce a net red-shift
in transition region lines. 

More recently 
\citet{2006ApJ...638.1086P} 
found that 3d numerical models that span the photosphere to corona produce red-shifts in
transition region lines. In these models coronal heating is due the Joule dissipation of currents
produced as the magnetic field is stressed and braided by photospheric motions, but the 
ultimate source of the red-shifts is not clear. 
\citet{2006ApJ...642..579S}  
used 1d models of coronal loops to show that red-shifts, and the observed
DEM at temperatures below $10^5$~K could be the result of transient heating near the loop 
footpoints.


In this work we extend the work of Peter {\rm et al.} and Spadaro {\rm et al.} to show that 
transition region red-shifts are naturally produced in episodically heated models where the
average volumetric heating scale height lies between that of the chromospheric pressure
scale height of 200~km and the coronal scale height of 50~Mm. 

\section{3d simulations of the solar atmosphere}

We solve the MHD equations using BIFROST \citep{Gudiksen+etal2010}. This code employs a high-order finite difference scheme and includes a high-order artificial viscosity in order to maintain numerical stability. These terms are also the source of magnetic and viscous heating. Non-grey, optically thick radiative losses including the effects of scattering are included, as are optically thin radiative losses for the upper chromosphere and corona and conduction along the field lines in the corona.

\subsection{Simulation description}

In this paper we will use results from four 
3d simulations of the solar atmosphere. These models span the region from 1.4~Mm below the
photosphere to 14~Mm above the photosphere, and thus include the upper convection zone, the
photosphere, chromosphere, transition region and the (lower) corona. Photospheric and lower
chromospheric radiative losses are balanced by an in-flowing heat flux at the bottom boundary.
The chromosphere and regions above are maintained by the dissipation of acoustic shock waves that are
generated in the convection zone and photosphere, but perhaps more importantly by the dissipation of the Poynting flux that is injected into the upper atmosphere. This Poynting flux arises as a result of the braiding of magnetic field lines as well as the injection of new ``emerging'' magnetic flux at the lower boundary. These processes have been described earlier in
the context of numerical simulations of the outer solar atmosphere 
\citep{2005ApJ...618.1020G,2004IAUS..223..385H,2005ESASP.592..483H,2007ASPC..368..107H}
and also including flux emergence 
\citep{2008ApJ...679..871M,2009ApJ...702..129M}. 

Three of the models are comprised of $512\times 512\times 325$ points spanning a region of
$16.6\times 16.6\times 15.5$~Mm$^3$. These models differ in their total unsigned magnetic
flux, the weak field model henceforth named ``A1'' has $\aver{\labs B_z\rabs}=20$~G in the photosphere.  On the other hand, the stronger field model, named ``A2'' has $\aver{\labs B_z\rabs}\approx 100$~G. In addition, in the stronger field model a flux sheet 
is injected at the bottom boundary. Finally, the ``A3'' model is initially identical to A2 but with very little flux injected at the bottom boundary. The A1 and A2 models have been run for 1~hour solar time while the A3 model for roughly $45$~minutes.

\begin{figure}
  \includegraphics[width=\columnwidth]{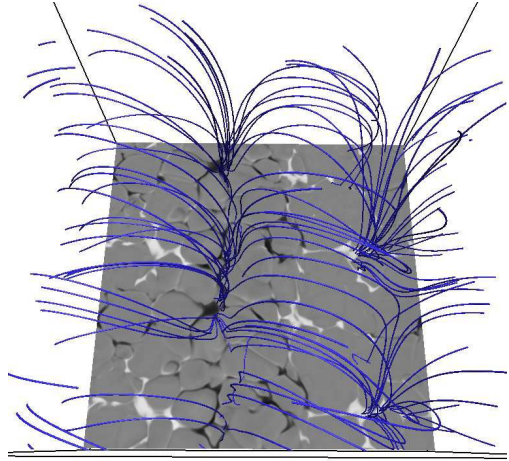}
  \caption{Vertical magnetic field $B_z$ in the photosphere and magnetic field lines that 
reach the corona (in blue). The image is from time $t=2700$~s in the A2 simulation.
  \label{fig:bzb_qsmag_8t_XX}}
\end{figure}

The fourth model, ``B1'', has lower resolution --- $256\times 128\times 160$ points --- and spans a smaller region of $16.6\times 8.3\times 15.5$~Mm$^3$. The unsigned magnetic field is of the same order, but slightly weaker $\aver{\labs B_z\rabs}\approx 75$~G, than in  as in the strong field model above. This low resolution model also includes an injected 
emerging flux sheet. The B1 model has been run for 1~hour solar time.

In figure~\ref{fig:bzb_qsmag_8t_XX} we show the vertical magnetic field in the photosphere along with a subset of field lines that reach coronal heights in the A2 simulation. This magnetic topology is typical for the simulations discussed here.


Also typical for these models is a temperature structure 
in the transition region and corona 
where the temperature begins to rise some $1.5$~Mm above the photosphere and has reached $1.0$~MK already at heights of $3-4$~Mm. However, the structure of the atmosphere is far from horizontally homogeneous and
we find very hot plasma ($>2$~MK for the A2 run) already at heights of $1.5$~Mm, while at the same time we find cool gas with temperatures $<10$~kK even at heights up to $6$~Mm. 




\subsection{Characteristics of simulated TR and coronal lines}

Let us now consider some examples of simulated spectral lines formed in the transition 
region and corona of these models. We have chosen lines that have been previously well observed with space borne instruments such as SOHO/SUMER or are currently observable with Hinode/EIS. These lines cover a wide span of temperature in through the outer solar atmosphere and are summarized in table~\ref{tab:lines}.

\begin{table}
\begin{tabular}{lrrr}
\hline
\\
Line ID & $\lambda$ [nm]  & $\log(T_{\rm max})$ & mass [amu] \\
\hline
\\
\HeII & $25.63$ & $4.9$ & $4.0$ \\
\CIV & $154.82$ & $5.0$ & $12.0$ \\
\OVI & $103.19$ & $5.5$ & $16.0$ \\
\SiVII & $27.54$ & $5.8$ & $28.1$ \\ 
\FeXII & $19.51$ & $6.1$ & $55.9$ \\
\FeXV & $28.42$ & $6.3$ &  $55.9$ \\
\\
\end{tabular}
\caption{\label{tab:lines} List of lines synthesized from the simulations described.}
\end{table}

The synthesized intensities $I_\nu$ in the various EUV spectral lines mentioned above are computed assuming that the lines are optically thin and that the ionization state is in equilibrium

\begin{equation}
I_\nu={h\nu\over 4\pi}\int_s\phi_\nu(u,T_{\rm g})n_{\rm e}n_{\rm H}g(T_{\rm g})ds
\end{equation}                  

\noindent where the integration is carried out along the line of sight, or in this case vertically 
through the box along the $z$ axis. The line profile is computed assuming Doppler broadening.





\begin{figure}
  \resizebox{0.48\hsize}{!}{\includegraphics[clip=true]{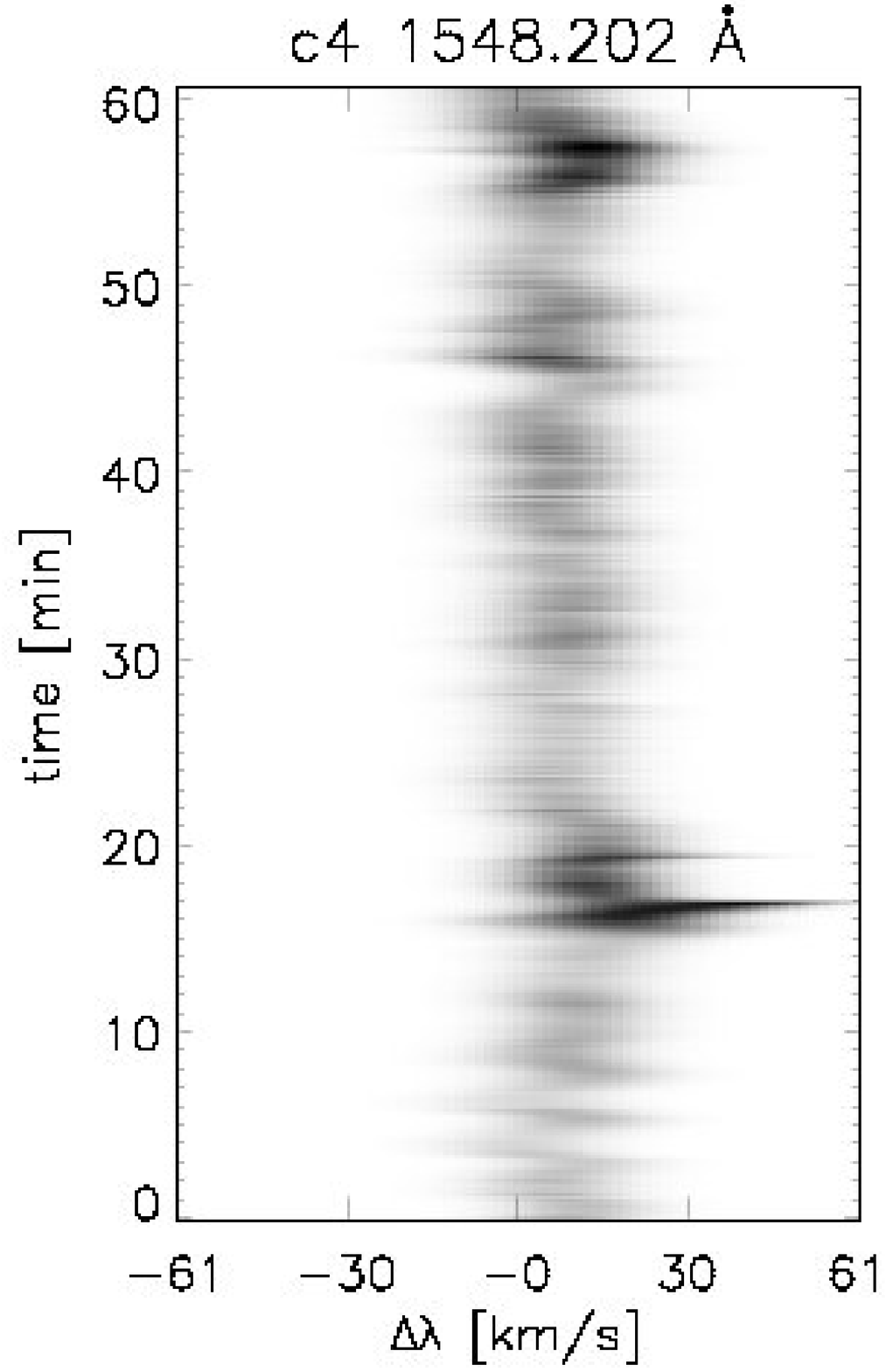}}
  \resizebox{0.48\hsize}{!}{\includegraphics[clip=true]{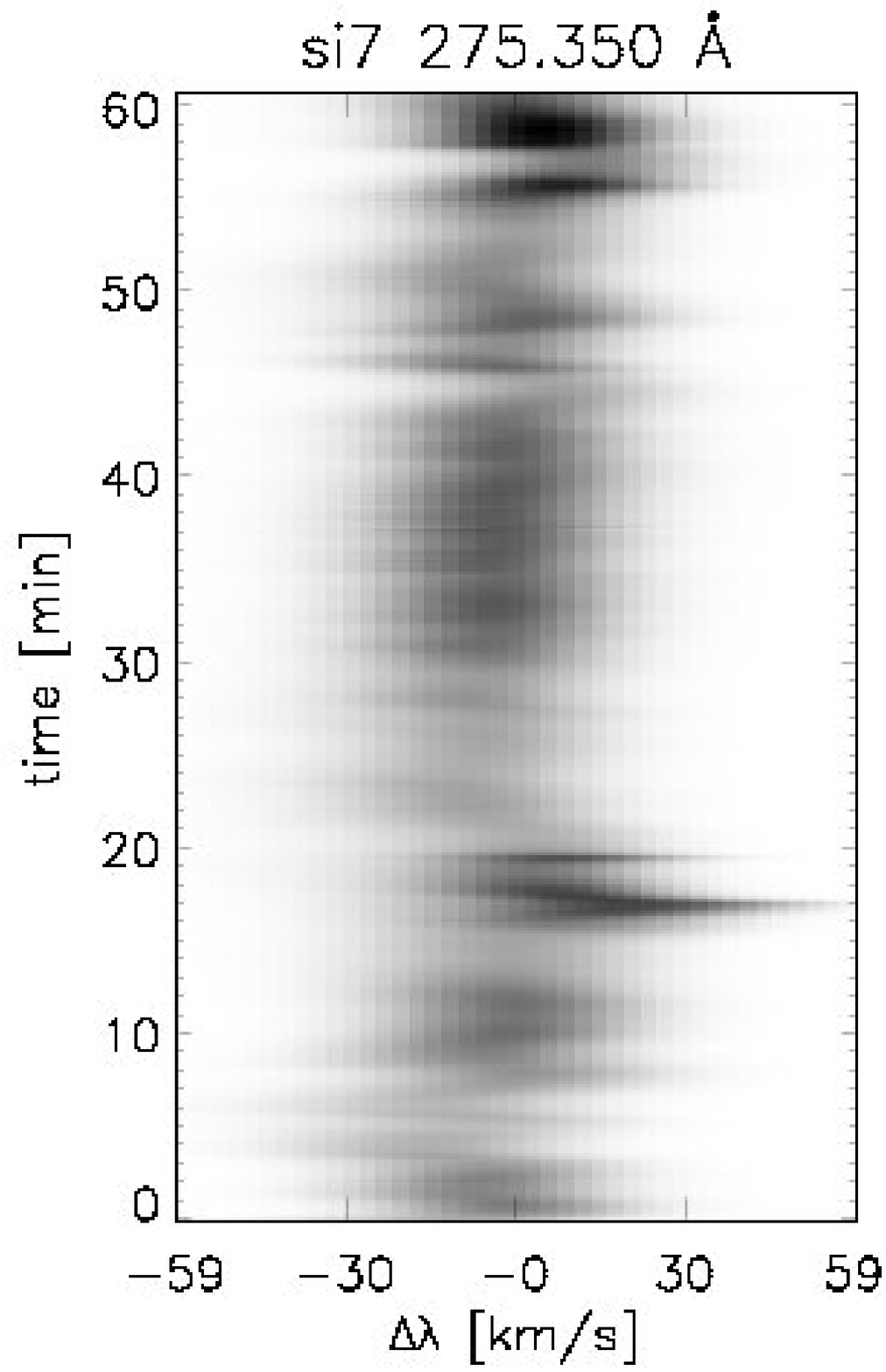}}
  \caption{Intensity profiles of the \CIV\ and \SiVII\ lines as functions of time from the low resolution simulation B1.
  \label{fig:tr_prof}}
\end{figure}

Examine now how the line profiles vary as a function of time at a given typical location 
picked more or less at random in the B1 simulation. The transition region line profiles, exemplified by \CIV\ and \SiVII\ representative of the lower and upper transition region respectively, are shown in figure~\ref{fig:tr_prof}. These lines show a fairly similar evolution with significant variations in line intensity, shift, and width on time-scales down to less than a minute. We see strong red-shift events accompanied by brightening such as at 18~min and at 56~min, and a weaker blue-shifted event at 47~min. It is clearly evident that the 
bulk of the \CIV\ line is red-shifted at almost all times, while the \SiVII\ line has little net shift.

\begin{figure} 
\resizebox{0.48\hsize}{!}{\includegraphics[clip=true]{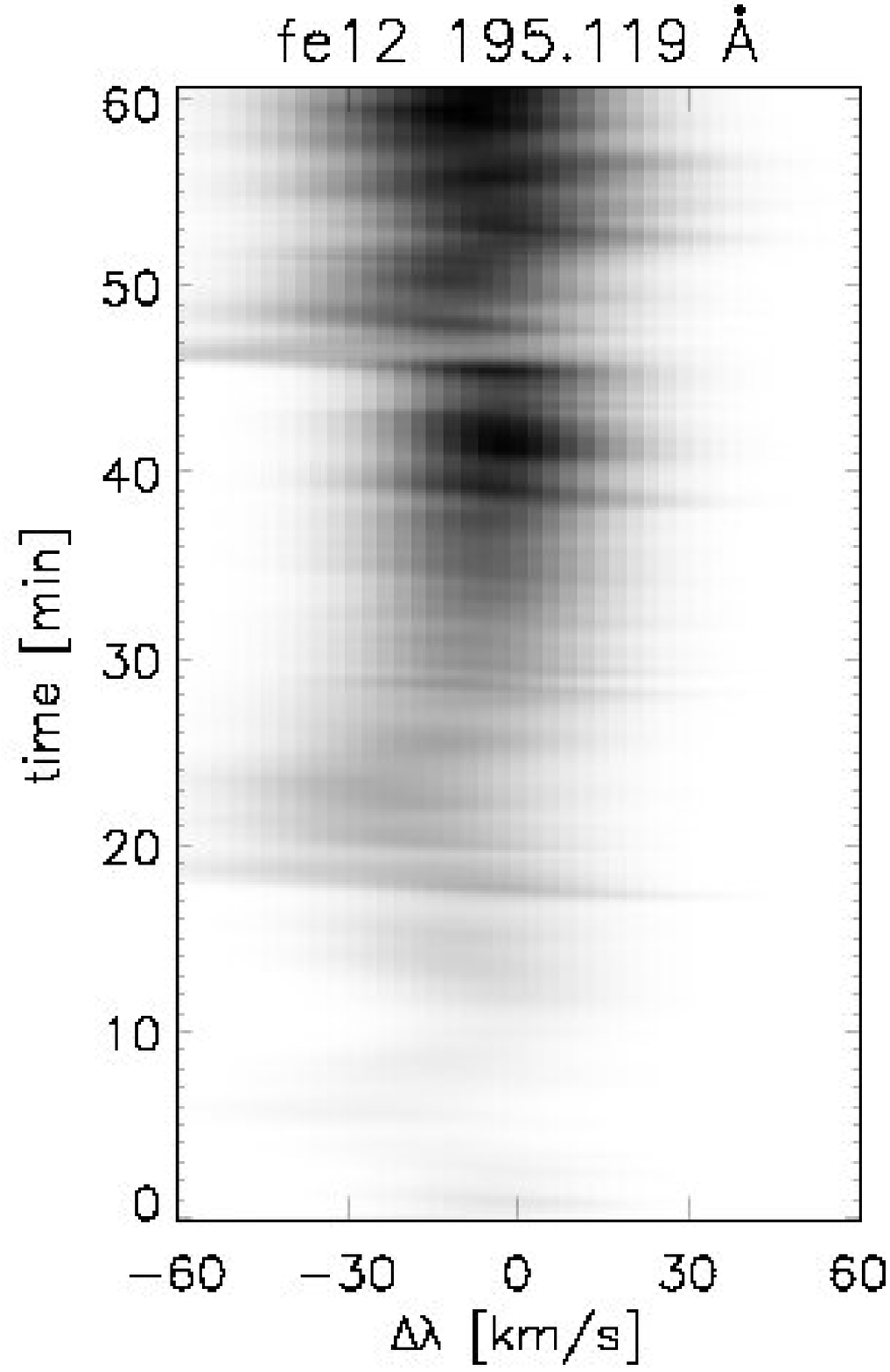}} \resizebox{0.48\hsize}{!}{\includegraphics[clip=true]{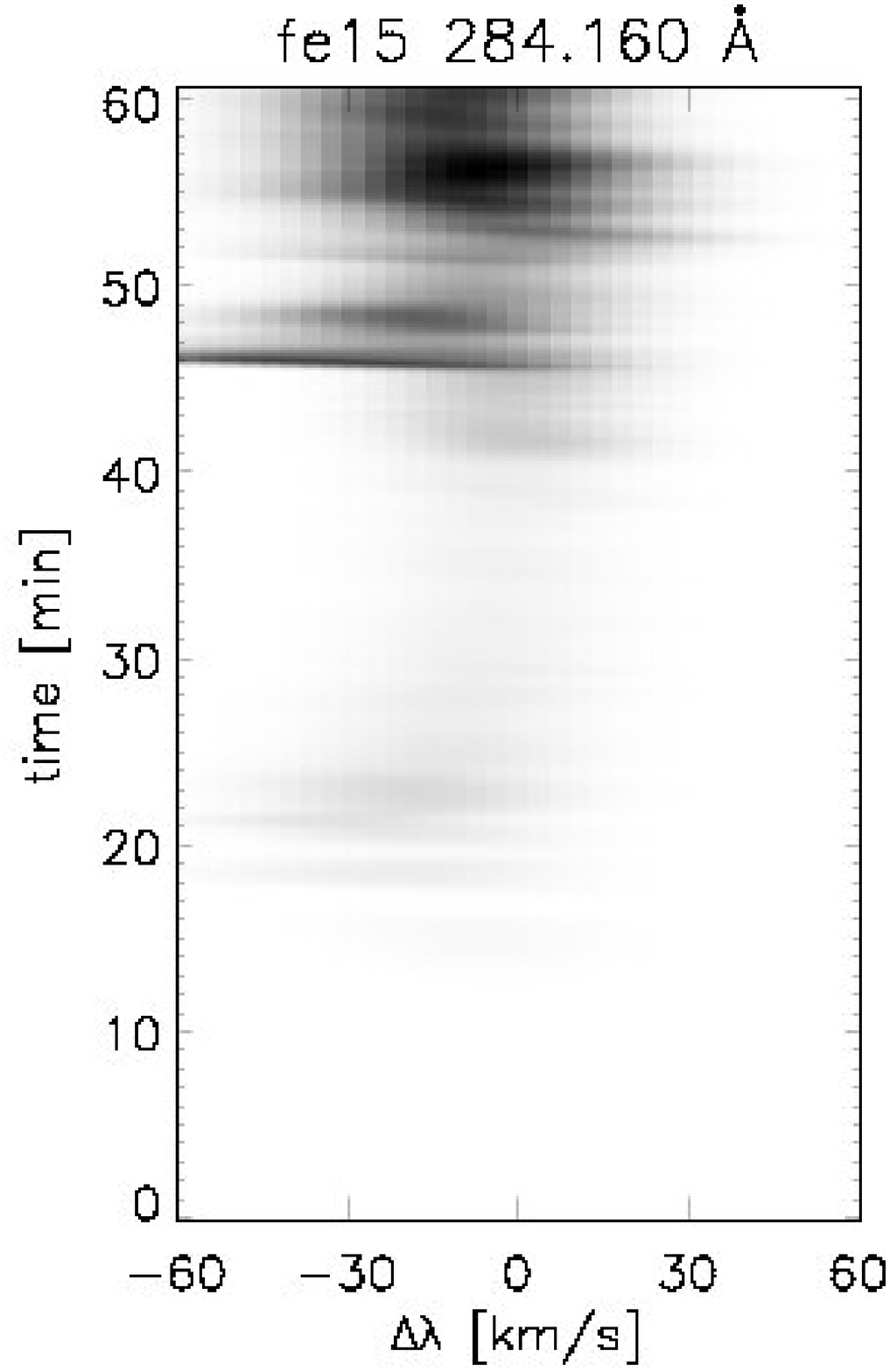}}
  \caption{Intensity profiles of the \FeXII\ and \FeXV\ lines as functions of time from the low resolution simulation B1.
  \label{fig:corona_prof}}
\end{figure}

The coronal line profiles, exemplified by \FeXII\ and \FeXV\ are shown in figure~\ref{fig:corona_prof}. Also here we find large variations in the line parameters on several timescales including those shorter than 1~min. The coronal lines become brighter as the corona grows hotter at times later than 20~min in the simulation. The strong brightening and blue-shift seen at 47~min is found to be associated with a strong heating event (the joule heating in the B1 model is shown in figure~\ref{fig:j2r_bz_trt}, where the event causing the line variation is clearly visible). Note that the coronal lines at this location are preferentially blue-shifted.

\begin{figure*}
\includegraphics[width=2\columnwidth]{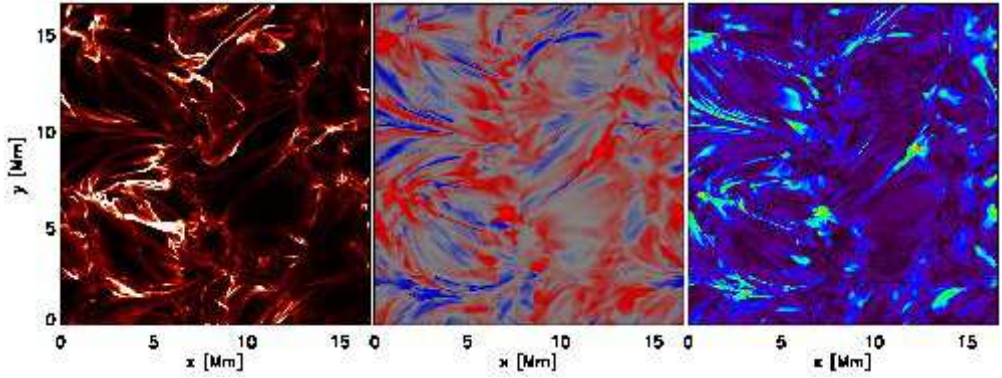}
  \caption{Total intensity, doppler shift, and line width 
in the \CIV\ line at time $t=2700$~s in the A2 simulation. The velocity scale is from $-40$~km/s (blue) to $40$~km/s (red). Line widths range from narrow black to wide yellow/red with a maximum of 51~km/s. The average line shift in the \CIV\ line is $6.6$~km/s.
  \label{fig:iuw_tr_map}}
\end{figure*}

\begin{figure*}
\includegraphics[width=2\columnwidth]{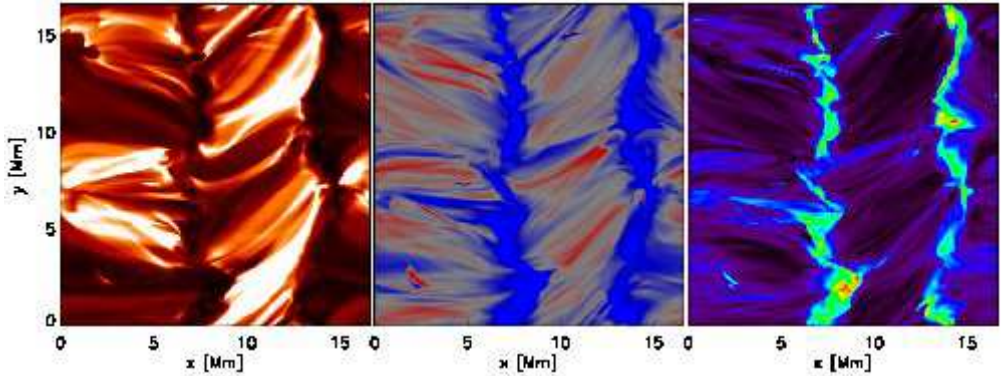}
  \caption{Total intensity, doppler shift, and line width in the \FeXII\ line at time $t=2700$~s in the A2 simulation. The velocity scale is from $-40$~km/s (blue) to $40$~km/s (red). Line widths range from narrow black to wide yellow/red with a maximum of 76~km/s. The average line shift in the \FeXII\ line is $-2.7$~km/s.
  \label{fig:iuw_cor_map}}
\end{figure*}

We can also examine the properties of the lines by examining the lower order moments; the total intensity, average line shift, and average line width.
In figure~\ref{fig:iuw_tr_map} we show the line moments of the lower transition region \CIV\ line at time $t=2700$~s in the A2 run. The left panel shows the total intensity, the middle panel the line shift, and the right panel the line width. 
While the intensities do not obviously reflect the topology of the magnetic field, the 
line shift and width are more clearly related to the field. Intense red- and blue-shifts are generally concentrated to positions where the field comes up vertically from the chromosphere below, {\it i.e} in the loop footpoints (see figure~\ref{fig:bzb_qsmag_8t_XX}). In the regions between the footpoints we also find both red- and blue-shifts aligned along the field lines that stretch from the footpoints concentrated in bands near $x=7$~Mm and near $x=13$~Mm. 
Similarly, we find that the greatest line widths are found in the footpoint bands, and the \CIV\ lines are in generally much narrower in the regions between. However, note that we also find some large line width regions strething in thin strings between the footpoint bands and aligned with the field lines shown in figure~\ref{fig:bzb_qsmag_8t_XX}.

The pattern of intensities seen in the coronal lines is quite different, as seen for example in the \FeXII\ line intensities shown in figure~\ref{fig:iuw_cor_map} which clearly delineate the magnetic field lines that stretch more or less horizontally between the footpoints in the form of loops. On the other hand, the structure of line shifts 
are very similar to the structure found in the transition region lines. Blue-shifts are generally larger (the maximum blue-shift is $-75$~km/s while the maximum red-shift is only $35$~km/s) and concentrated to the loop footpoints. Notice that footpoint intensities are much lower than in the bodies of the loops between the footpoints, presumably as footpoint temperatures are too low to give emission in coronal lines. 

\section{Coronal heating in this model}

Let us now put the ``observed'' line profiles described into context with the coronal heating that characterizes these models.

\begin{figure}
\resizebox{\hsize}{!}{\includegraphics[clip=true]{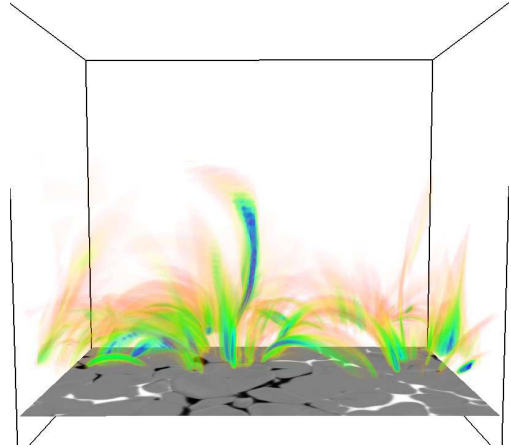}}    
  \caption{Current density squared per particle ($\eta j^2/\rho$) at time $t=2810$~s in simulation B1 (red weakest, blue strongest) and vertical magnetic field $B_z$ in the photosphere. 
  \label{fig:j2r_bz_trt}}
\end{figure}

The heating per particle ($\propto j^2/\rho$) is concentrated to certain regions and structures that, in general, follow the magnetic field lines as shown in figure~\ref{fig:j2r_bz_trt} where the heating in the B1 simulation at time $2810$~s. This heating is highly irregular both in space and time but is in large part concentrated to the transition region and lower corona. Note that regions of high heating (colored in green --- blue) in part are aligned in loop like structures, in part in linear structures that stretch high into the corona. In addition, we find a weaker  component (colored in red) which, though it also follows the structure of the magnetic field, seems much more evenly spread out and space filling than the regions of (sporadic) high heating. In time, the linear regions of high heating are short lived, of order
$100$~s or shorter. The linear structures sometimes also move horizontally over a thousand kilometers or so during their lifetime. The loop shaped structures are longer lived, though the heating varies continuously in time also here. 

Thus, high heating events occur in spatially localized bursts that lead to increases in the 
temperature but also large velocities. For example, the large linear structure that reaches high into the corona, seen in the middle of the left panel of figure~\ref{fig:j2r_bz_trt} is the event that is visible in figure~\ref{fig:tr_prof} as a line shift of the transition region lines at time $t=47$~minutes and that is also evident in figure~\ref{fig:corona_prof} as a line shift and, especially for the \FeXV\ line, by an increase in the intensity.

\begin{figure}
  \includegraphics[width=\columnwidth]{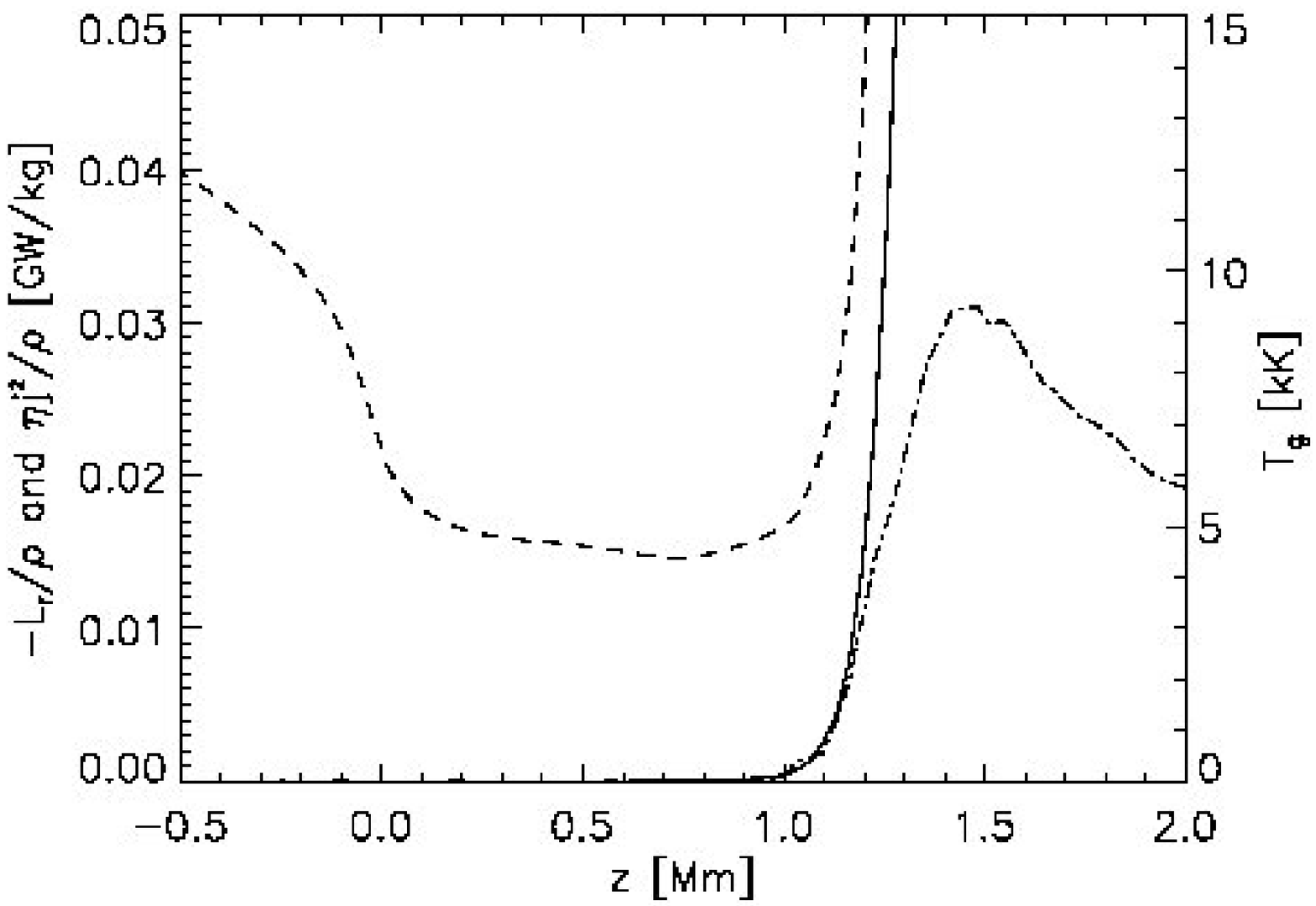}
  \includegraphics[width=\columnwidth]{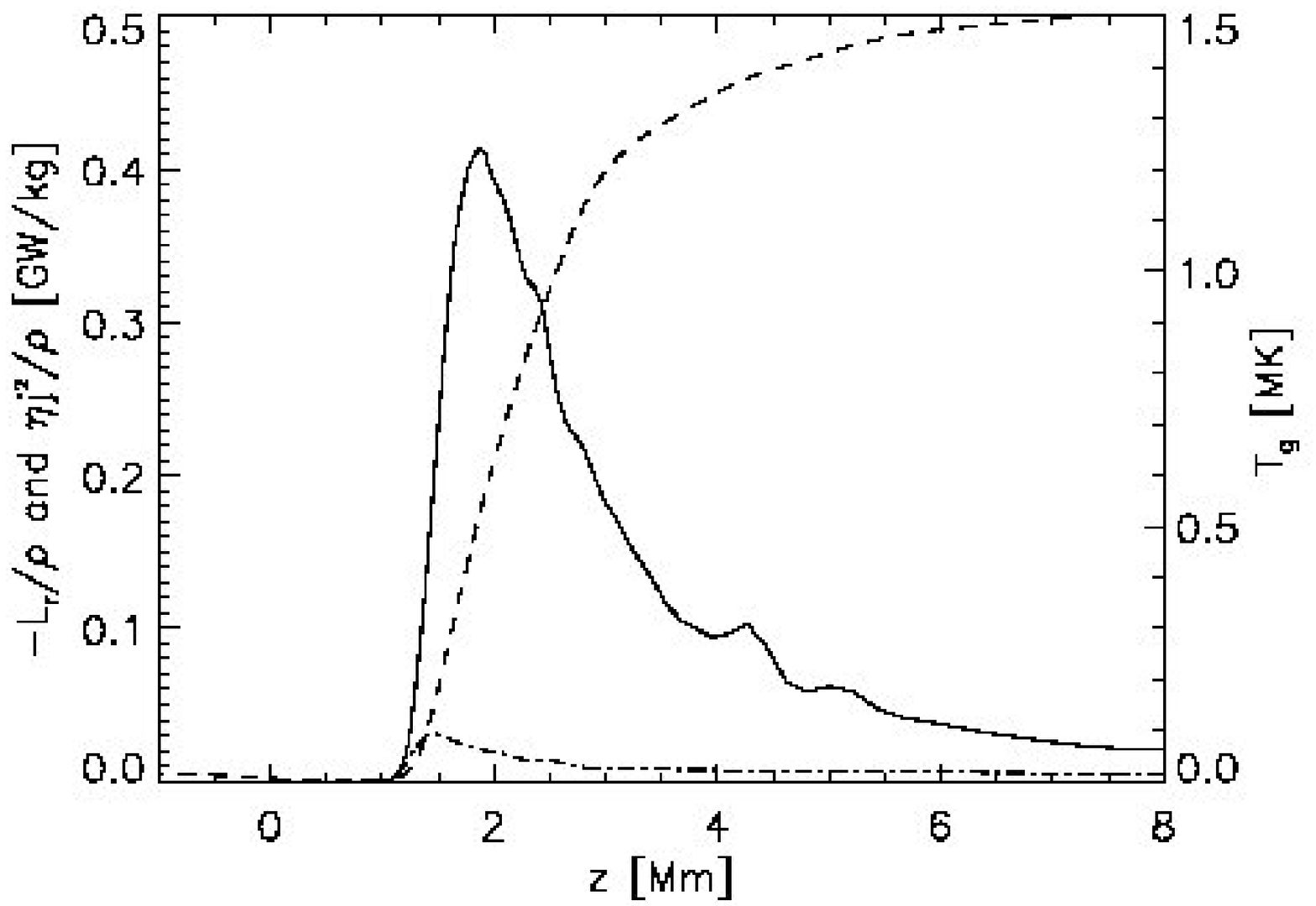}
  \caption{Average joule heating $\eta j^2/\rho$ (solid line) and average radiative losses $L_r=-n_{\rm e}n_{\rm H}f(T)/\rho$ (dot dashed line) per unit mass in the chromosphere and lower transition region (upper panel) and corona in simulation B1. The average temperature is over plotted for reference (dashed line).
  \label{fig:avg_j2r_tg}}
\end{figure}

The concentration of heating towards the transition region and lower corona can be understood by considering the various scale heights involved. We find that joule heating caused by the braiding of the magnetic field decreases exponentially with height from the upper photosphere with a scale height that is closely related to the scale height of the magnetic energy density ($B^2/2\mu_0$). This is the same result as found earlier in the \citet{2005ApJ...618.1020G} 
model (see figure 7 of that paper), where a heating scale height of order 650~km was found in the chromosphere, rising to 2.5~Mm in the low corona and 5~Mm in the extended corona. Similar heating scale heights are found in 
\citet{2008ApJ...679..871M,2009ApJ...702..129M}. 
They are much greater than the chromospheric pressure scale height of 200~km while, on the other hand, being  much smaller than the transition region scale height of 6~Mm or the 60~Mm coronal scale height. Thus, in models where chromospheric/coronal heating is determined by photospheric work on the magnetic field --- with or without flux emergence --- we find that the heating per unit mass (or equivalently per particle) necessarily is concentrated towards the transition region and the low corona: Low in the chromosphere the particle density is very high and therefore the heating rate per particle low. At the same time radiative losses are very efficient at these heights, thus the plasma easily rids itself of deposited energy. However, since the scale height of heating ($\sim 650$~km) is greater than that of the particle density ($\sim 200$~km) in the roughly isothermal chromosphere, the heating per particle rises exponentially with height. At some point the heating overwhelms the plasma's ability to radiate, and the temperature rises to coronal temperatures, as illustrated in figures~\ref{fig:avg_j2r_tg}. This figure shows the joule heating ($\eta j^2$) and radiative cooling ($L_r=-n_{\rm e}en_{\rm H}f(T)$) per unit mass as a function of height.  The temperature continues to rise until thermal conduction becomes effective enough to balance the deposited energy at temperatures approaching 1~MK. This temperature is high enough that the density scale height becomes much larger than the heating scale height and the heating per particle therefore decreases exponentially, roughly as the heating rate (or the magnetic energy density) itself. 

\section{Discussion \& Conclusions}

Thus simulations of this sort lead to a coronal heating model where sporadic heating of material occurs mainly in the vicinity of the upper chromosphere, transition region, and lower corona. When a heating event occurs in the upper chromosphere or transition region cooler material is rapidly heated to coronal temperatures. When heating occurs in the lower corona the thermal conduction increases rapidly with rising coronal temperatures and the cooler chromospheric material just below can be heated by a thermal wave.  It is our contention that these processes happen sufficiently rapidly that material is heated more or less ``in place'' before it has time to move. Thus, newly heated transition material will be characterized by higher than hydrostatic  pressures, and we expect, {\it on average}, to find a plug of high pressure material that seeks to expand into regions of lower pressure on either side; downwards towards the chromosphere and upwards towards the corona.

\begin{figure}
  \includegraphics[width=\columnwidth]{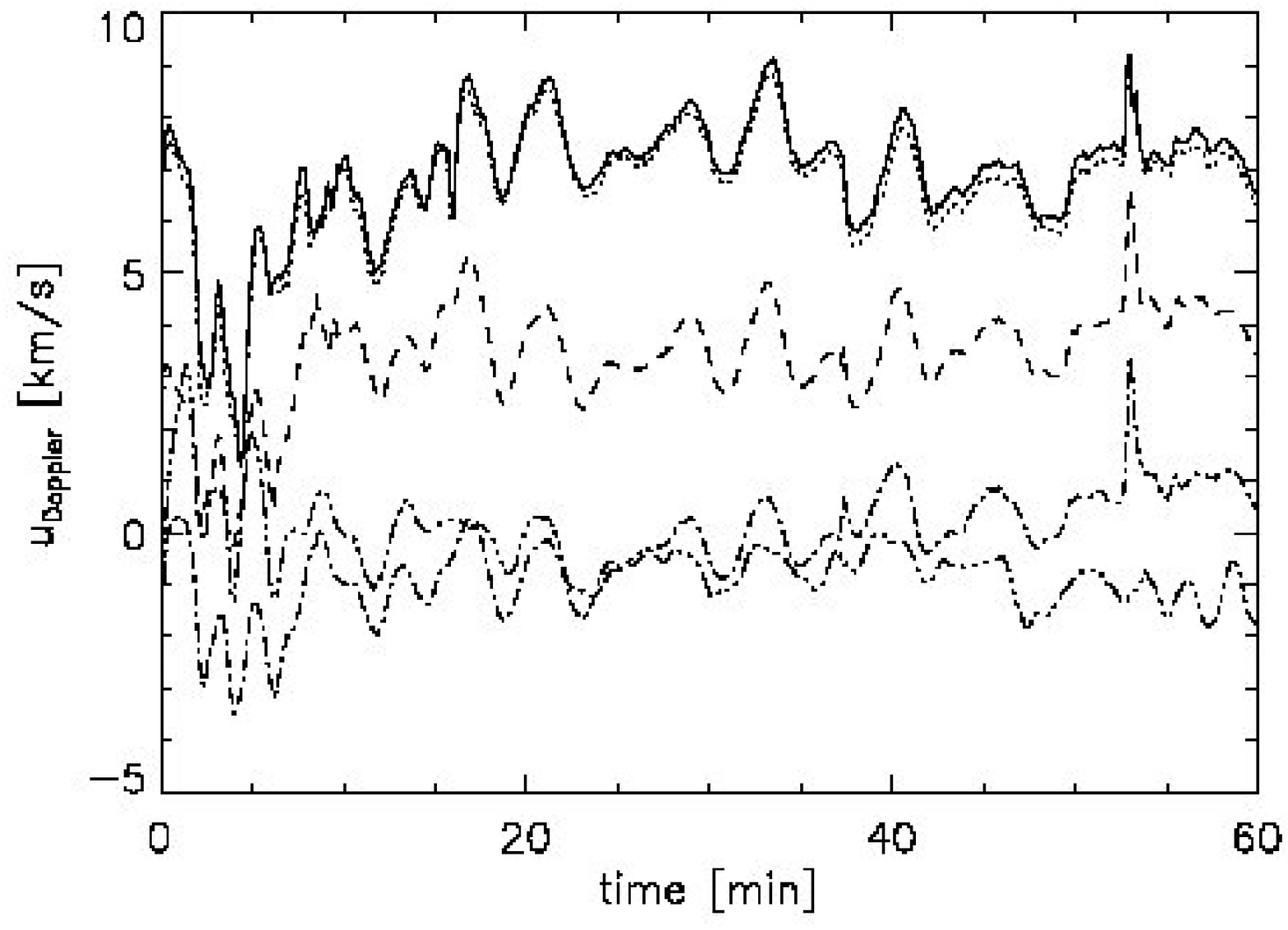}
  \includegraphics[width=\columnwidth]{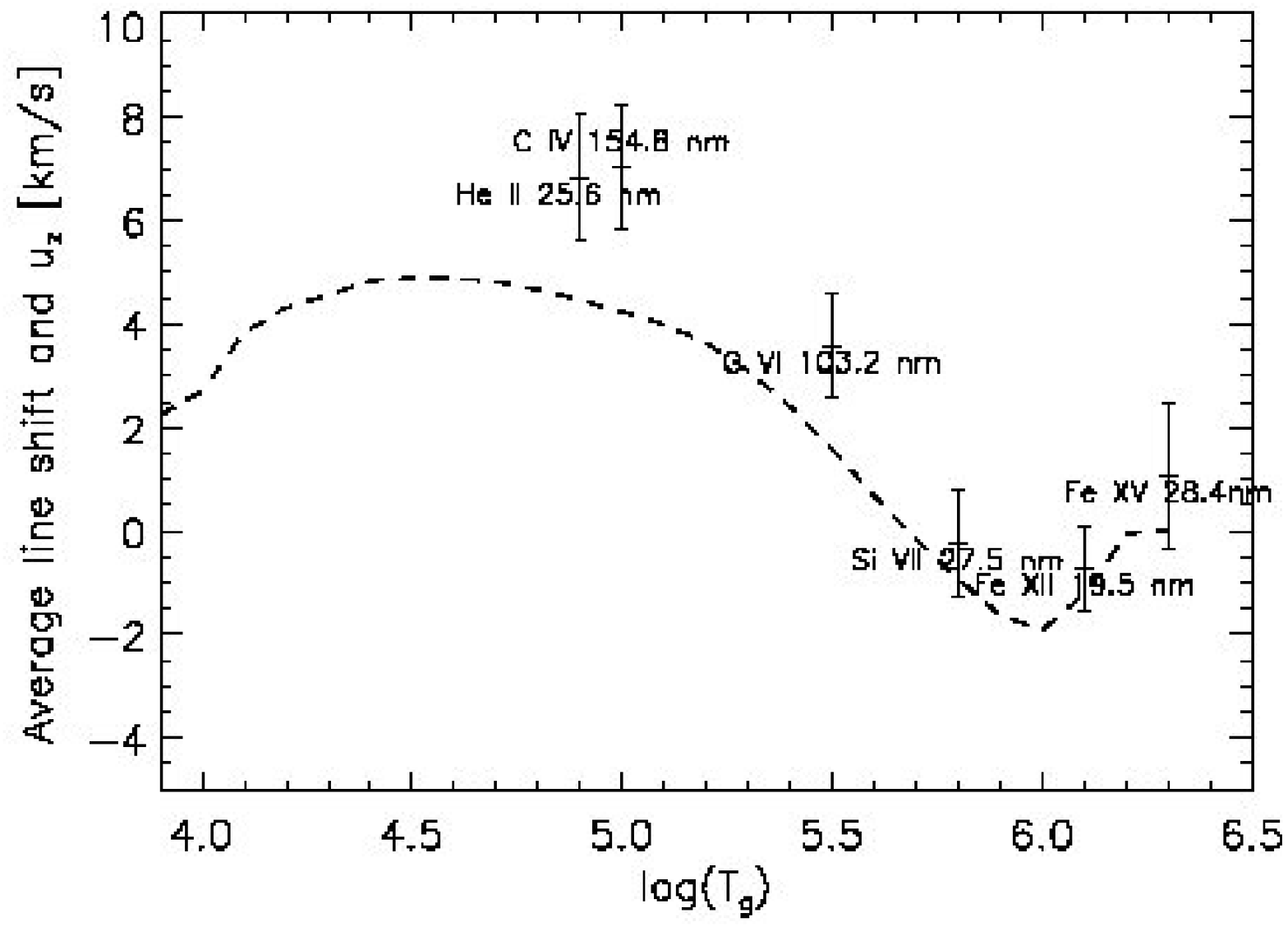}
  \caption{Average line shift in the B1 model as a function of time for the lines of table~\ref{tab:lines} (upper panel): \HeII\ (dotted), \CIV\ (solid), \OVI\ (dash), \SiVII\ (dash dot), and \FeXII\ (dash dot dot dot). The lower panel shows the average line shifts
and the average velocity (dashed) as a function of log temperature in simulation B1. The error bars in the average line shift are given by the standard deviation in the average velocities as a function of time.
  \label{fig:redshift}}
\end{figure}

The flows away from the high pressure plug are reflected in the average velocities found in the transition region and lower corona as shown in figure~\ref{fig:redshift}. Average line shifts found in simulation B1 are relatively constant, showing a large apparent downflow in the transition region and a smaller apparent upflow at coronal temperatures. These line shifts represent an actual material flow, on average material is flowing away from the temperature of some $7\times 10^5$~K with a maximum downward velocity of some $5$~km/s at $3\times 10^4$~K and a maximum upward velocity of $2$~km/s at 1~MK as shown in the lower panel of figure~\ref{fig:redshift}. The line shifts reflect this flow trend fairly accurately but due to the weighting of the line profile with density ($\propto n_{\rm e}n_{\rm H}$ the line shifts are exaggerated in the transition region. We find that the {\it average} line shift in the coolest transition region lines, \HeII\ and \CIV\, are of order $7-8$~km/s. The \OVI\ line is also shifted, but less, with an average that lies near $4$~km/s. The upper transition region \SiVII\ line is initially blue-shifted, but as time passes in the simulation the line shift  decreases towards zero and is slightly red-shifted at the end of the run. Finally the \FeXII\ line has an average line shift of zero initially but is slightly blue-shifted at the end of the simulation run.

Note that though a net red-shift is observed in transition region lines which corresponds to a material flow out of the transition region, the coronal mass is actually constant or growing in these simulations. Thus, material does not {\it flow} into the corona through the transition region. Rather, more material is heated rapidly to high temperatures than can be supported, and this material is subsequently pushed out of the upper transition region into the corona and chromosphere. 

\begin{figure}
  \includegraphics[width=\columnwidth]{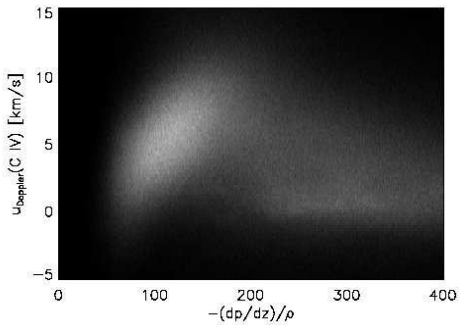}
  \caption{Doppler shift of the \CIV line versus the maximum downward oriented
pressure gradient  $({dp/dz})/\rho$ along the atmospheric column where the Doppler 
shift was measured for the entire model run of simulation B1 (see text for details).
  \label{fig:dpdz_udopp_corr}}
\end{figure}

Further support for this scenario dominating the computed models is shown in figure~\ref{fig:dpdz_udopp_corr} where we plot
the correlation between the red-shift measured in the \CIV\ transition region line and the maximum downward oriented pressure gradient $({dp/dz})/\rho$. The figure shows
that high red-shift occurs in regions with a large downward directed pressure gradient. Note that figure~\ref{fig:dpdz_udopp_corr} also shows that a fairly large population of large downward directed pressure gradients co-exist with line shifts that are close to zero. This population is representative of points where the magnetic field is mainly horizontal; {\it ie} where there is a significant contribution from the Lorentz force, presumably located in the body of low lying cooler loops. The latter population does not have much impact on the average line shifts.

In summary, we find that outflows from the upper transition region can be explained as a natural consequence of episodically heated models where the heating per particle is concentrated towards the lower corona. Chromospheric material is heated {\it in place} to coronal temperatures, or rapidly heated coronal material just above the transition region leads to a thermal conduction front heating the chromospheric material just below the joule dissipation region to coronal temperatures; both variations lead to a high pressure plug of material at upper transition region temperatures that relaxes towards equilibrium by expelling material. We have further shown that heating with these characteristics naturally arises when coronal heating is assumed to be caused by the dissipation resulting from the braiding of the photospheric magnetic field. This mechanism leads to joule heating that has an exponential decrease with height closely related to the scale height in the magnetic energy density.

\begin{acknowledgements}
V.H.H. thanks the NAOJ for hospitality during the summer of 2009.
This reasearch was supported the European Commission funded
Research Training Network SOLAIRE. 
It was also supported from grants of computing time from the Norwegian Programme for Supercomputing from the Norwegian Research Council Hinode grant.
\end{acknowledgements}

\bibliographystyle{aa}

\end{document}